\documentclass[preprint,showpacs,preprintnumbers,amsmath,amssymb,
nofootinbib,eqsecnum,12pt,floatfix]{revtex4}

\usepackage{graphicx}

\newlength{\dummysp}
\newcommand{\half}{\frac{1}{2}}

\newcommand{\beq}{\begin{eqnarray}}
\newcommand{\eeq}{\end{eqnarray}}
\newcommand{\nnn}{ \nonumber \\ }
\newcommand{\ddd}{ \nnn && }
\newcommand{\p}{{\partial}}

\newcommand{\gappeq}{\mathrel{\rlap {\raise.5ex\hbox{$>$}}
{\lower.5ex\hbox{$\sim$}}}}
\newcommand{\lappeq}{\mathrel{\rlap{\raise.5ex\hbox{$<$}}
{\lower.5ex\hbox{$\sim$}}}}
\newcommand{\myref}[1]{(\ref{#1})}

\newcommand{\ben}{\begin{enumerate}}
\newcommand{\een}{\end{enumerate}}
\newcommand{\bit}{\begin{itemize}}
\newcommand{\eit}{\end{itemize}}

\newcommand{\Ncal}{{\cal N}}

\newcommand{\Rhat}{{\hat R}}
\newcommand{\chitil}{{\tilde \chi}}
\newcommand{\gbar}{{\bar g}}

\def\[{\left [}
\def\]{\right ]}
\def\({\left (}
\def\){\right )}

\begin{document}

\title{Backward running or absence of running from Creutz ratios}

\author{Joel Giedt}
\email{giedtj@rpi.edu}
\author{Evan Weinberg}
\email{weinbe2@rpi.edu}
\affiliation{Department of Physics, Applied Physics and Astronomy,
Rensselaer Polytechnic Institute, 110 8th Street, Troy NY 12065 USA}

\date{Aug.~19, 2011}

\begin{abstract}
We extract the running coupling based
on Creutz ratios in SU(2) lattice gauge theory
with two Dirac fermions in the adjoint representation.
Depending on how the extrapolation to zero fermion
mass is performed, either backward running or
an absence of running is observed at strong bare coupling.
This behavior is consistent with other
findings which indicate that this theory has
an infrared fixed point.
\end{abstract}



\maketitle

\section{Motivation}
Understanding the allowed model space in extensions to Standard Model 
is a research program that gains motivation at the onset of the Large Hadron Collider.
In technicolor models, the Higgs mechanism occurs through
condensation of new fermions that are subject to a gauge interaction
that is strong at the TeV scale \cite{Susskind:1978ms,Weinberg:1979bn}.  Walking technicolor is a version
of this theory that can address certain difficulties associated
with flavor-changing neutral currents \cite{Holdom:1981rm,
Holdom:1984sk,Yamawaki:1985zg,Bando:1986bg,Appelquist:1986an,
Appelquist:1986tr,Appelquist:1987fc}.  Higher representations
of the gauge group are believed to avoid problems with the S-parameter,
i.e.~electroweak precision constraints \cite{Eichten:1979ah,Lane:1989ej}.  All of this has motivated
the study of Minimal Walking Technicolor \cite{Sannino:2004qp} using lattice techniques,
in order to see where this theory lies with respect to the conformal
window.  In other words, there is no guarantee that the theory
actually walks; finding out one way or another is to solve a strong
coupling problem.  Here we discuss a method, based on the behavior
of Creutz ratios, that does just that.
The approach taken here is different from the Schr\"odinger functional
method that was employed for SU(3) gauge group with triplet fermions 
in \cite{Appelquist:2007hu}, SU(3) gauge group with sextet fermions
\cite{Shamir:2008pb}, and Minimal Walking Technicolor 
\cite{Hietanen:2009az,Bursa:2009tj,DeGrand:2011qd}.  The latter
works indicate the existence of an infrared fixed point.  We
obtain results consistent with that conclusion by this alternate 
Creutz ratio approach.

\section{The method}
In \cite{Bilgici:2008mt} 
a method for measuring the running gauge coupling on the lattice
using Creutz ratios was proposed.  In \cite{Bilgici:2009kh,Fodor:2009rb} this method was
applied to pure SU(3) lattice Yang-Mills,
and in \cite{Fodor:2009rb,Fodor:2009wk} this method was applied in
a preliminary way to SU(3) gauge group with
sixteen triplet fermions, and more recently in \cite{Ohki:2010sr}
to SU(2) gauge group with eight flavors of fundamental representation
fermions.  We will use this approach to
search for the phenomenon of backward running,
which is a smoking gun for a nontrivial infrared
fixed point, since forward running is assured
at weak coupling by perturbative methods.  We apply the method of Creutz ratios to the
theory of Minimal Walking Technicolor:  SU(2) gauge theory with two flavors
of fermions in the adjoint representation.
Our lattice action is Wilson fermions and a plaquette gauge action.

\subsection{Outline}
The proposal \cite{Bilgici:2008mt} defines the running 
coupling $g(L)$ at a scale $L=Na$ associated
with the spatial extent of the lattice, where
$N$ is the number of sites in spatial directions and
$a$ is the lattice spacing.\footnote{In our analysis we
take temporal extent $T=2L$, to allow for
accurate measurement of the partially conserved
axial current (PCAC) mass and reuse of the lattice
configurations for studies of the spectrum.  In fact,
this is an advantage of the Creutz ratio method over
Schr\"odinger functional studies.}  This is done through
the Creutz ratio \cite{Creutz:1980wj}:
\beq
\chi(I,J) = -\ln \frac{W(I,J) W(I-1,J-1)}{W(I,J-1) W(I-1,J)}
\eeq
where $W(I,J)$ is the expectation value
of the trace of the rectangular $I \times J$ Wilson
loop on the lattice.  Throughout, we only consider $\chi(I,I)$.  The Creutz ratio
is interpolated between values of $I$
to define a related function $\chitil(\Rhat)$ with
$\Rhat \equiv R/a$ continuous.  The interpolation requires a
matching of values at the points where $\Rhat$ is
half-integral:
\beq
\chitil(\Rhat) \equiv \chi(\Rhat+\half,\Rhat+\half)
= -\ln \frac{W(\Rhat+\half,\Rhat+\half) W(\Rhat-\half,\Rhat-\half)}
{W(\Rhat+\half,\Rhat-\half)^2}
\eeq
the logic being that the Wilson loops appearing
in the Creutz ratio have average size $R$.
In fact, taking the classical continuum limit,
one finds that the Creutz ratio is a finite difference approximation:
\beq
\Rhat^2 \chitil(\Rhat) \approx - RT \frac{\p^2}{\p R \p T} \ln W_{\text{cont.}}(R,T) \bigg|_{T=R},
\label{fund}
\eeq
where $W_{\text{cont.}}(R,T) \approx W(\Rhat,\hat T)$ is the Wilson loop in the
continuum language.  $T$ here should not be confused the temporal extent of
the lattice.

In the Creutz ratio method for determining $g(L)$, we choose
\beq
r = R/L= \text{fixed}.
\eeq
Note also that we have to deal with a fermion mass $m_q$, which we
take to be the PCAC mass.  Our approach will be to measure $g(L)$
at nonzero $m_q$ and then extrapolate to the $m_q=0$ limit.

The important thing that Bilgici et al.~have
noted is that at one loop in lattice perturbation
theory, including the effect of bosonic zero modes,
\beq
\Rhat^2 \chitil_{\text{1-loop}}(L,a|\Rhat) = k(r,N) g_0^2, \quad N=L/a
\eeq
which defines the quantity $k(r,N)$.  Here, $g_0$ is the bare coupling. It was found
in \cite{Bilgici:2008mt,Bilgici:2009kh} that in the limit of large $N$,
$k(r,N) \to k(r)$, so that it is an $L$ and $a$ independent
quantity in the continuum limit.  Thus we write $k(r)$ in what follows,
and will define the running coupling in terms of this $N \to \infty$ normalization
factor.  Fermions enter $\chitil$ at two-loop order, so $k$ does not depend
on $m_q$, since it is defined by the one-loop expression.

Thus we can make an $L$-dependent nonperturbative definition
of the running coupling $\bar g$ using the value of $\chitil(R)$
obtained from a simulation with bare coupling $g_0$:\footnote{Often
below we will use the lattice coupling $\beta=4/g_0^2$.}
\beq
\gbar^2(L) \equiv \lim_{a \to 0} \frac{1}{k(r)} \Rhat^2 \chitil(L,a|\Rhat), \quad
\Rhat a / L = \Rhat/N = r = \text{fixed}
\eeq
Note that as we shrink $a$, in order to hold $L$ fixed the
number of lattice sites $N$ increases, and so must the
size of the Wilson loops $\Rhat$.  Since these quantities
fall off exponentially with increasing $\Rhat$ (relative to noise) and larger
lattices are significantly more expensive to simulate, this
becomes a demanding computation, particularly with
dynamical fermions.

To get a handle on $\gbar^2(L)$,
we can study the scaling of the function:
\beq
g^2 (N, r, L) \equiv \frac{1}{k(r)} \Rhat^2 \chitil(N,a|\Rhat)
\bigg|_{\Rhat = r N}
\label{juju}
\eeq
On the right-hand side, the nonperturbative $\chitil(N,a|\Rhat)$
is evaluated with system size $L$.
Clearly if $L$ is held fixed while $N$ is increased,
the lattice spacing is decreasing toward the continuum limit.
Thus $g^2(N, r, L)$ of \myref{juju} defines the running coupling in a particular
(nonperturbative) scheme:
\beq
\gbar^2(r|L) \equiv \lim_{N \to \infty} g^2 (N, r, L).
\label{gsq}
\eeq
The constant $r$ is part of the renormalization
scheme.

A step-scaling analysis is used to follow the 
running of the coupling $\bar g$ with the scale $L$.
We will not actually use step-scaling in order
to search for the presence of an infrared fixed point.  Rather,
there is a simpler, qualitative behavior that
we are looking for, as we now describe.

\subsection{Expectations}
\label{expect}
\begin{figure}
\begin{center}
\includegraphics[width=3in,height=5in,angle=90]{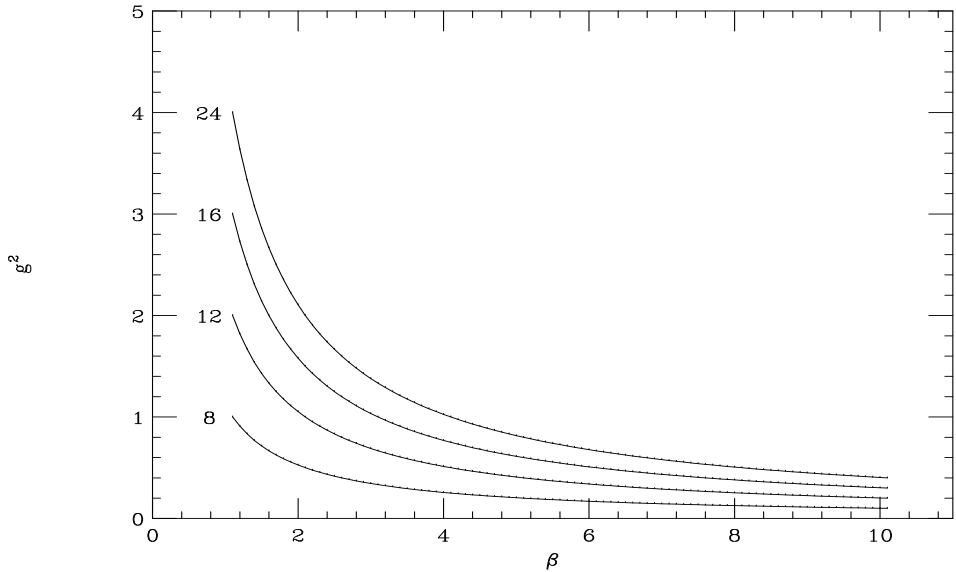}
\caption{Sketch of what should happen to $g^2(N,r,L)$ in a confining theory.
The lines are labeled with values of $N$.  \label{confine}}
\end{center}
\end{figure}

In a confining theory, $g^2(xN,r,xL) > g^2(N,r,L)$ for $x>1$.
This type of relationship means that if we hold $\beta$
fixed, the measured value of $g^2(N,r,L)$ increases
as $N$ increases, since $L=Na$ will increase proportionately.  
This leads to a behavior sketched in Fig.~\ref{confine}.
In fact, it is easy to convince oneself that a step-scaling
analysis applied using the curves in Fig.~\ref{confine}
will lead to a $g^2(L)$ that increases with $L$.

\begin{figure}
\begin{center}
\includegraphics[width=3in,height=5in,angle=90]{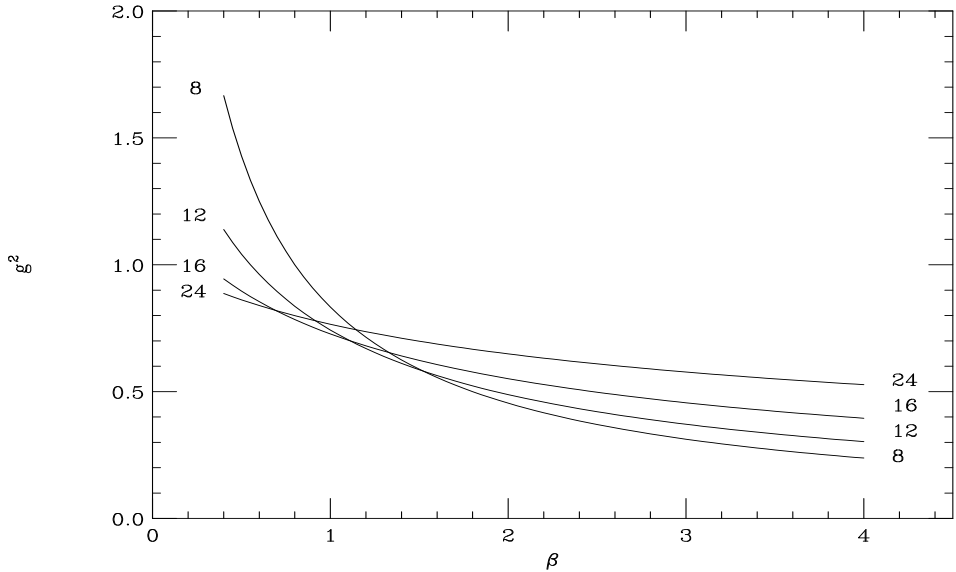}
\caption{Sketch of what should happen to $g^2(N,r,L)$ in a theory 
with an infrared fixed point. The lines are labeled with values of $N$. \label{fixed}}
\end{center}
\end{figure}

On the other hand, if there is an infrared fixed point,
the ordering of curves with increasing $N$ reverses once the bare coupling
$\beta$ goes past the fixed
point ($\beta < \beta_*(N)$), and $g^2(xN,r,xL) < g^2(N,r,L)$ occurs for $x>1$.
Here we note that the fixed point $\beta_*$ depends on $N$, due to lattice artifacts.
Indeed, such a ``backwards running'' was seen
at strong couplings in the SU(3) theory with
sixteen fundamental flavors in the work \cite{Fodor:2009wk,Fodor:2009rb}.
We expect something like this to also occur in the
present theory at strong coupling, under the assumption
that there is an infrared fixed point.
Thus we expect to obtain a behavior like the one sketched in Fig.~\ref{fixed}.
Note that the curves do not all cross at the same point, which is
why $\beta_*$ is a function of $N$.
Notice that there is a region of small $\beta$ where the ordering
of curves has clearly reversed, and that if we did a step scaling analysis
in this region, the flow would be backwards.

From the vantage point of the two sketches that we
have just presented, it is clear that a full blown
step scaling analysis is not needed in order to
detect the presence of an infrared fixed point.  All that is needed
is a plot of $g^2(N,r,L)$ versus $\beta$ for the various $N$.
In order to decide whether or not there is an infrared fixed point,
the question that we need to answer is just this:
``Which of the two figures does the plotted data look like,
Fig.~\ref{confine} or Fig.~\ref{fixed}?''
In fact even less is needed:  it is already known from
perturbation theory at weak coupling that the curves are
ordered with respect to $L$ as shown at large $\beta$.  So what we need
is just to show the reversal of ordering with respect to $L$ at small $\beta$.
This will be the focus of our analysis.

\section{Analysis}

\subsection{Smearing}
We follow Bilgici et al.~and apply APE smearing \cite{Albanese:1987ds,Teper:1987wt} for the
links that are used in the calculation of Wilson loops:  
\beq
U_\mu(x) &\to& P \{ U_\mu(x) + \alpha \sum_{\nu \not= \mu} [
U_\nu(x) U_\mu(x + a \hat\nu) U_\nu^\dagger(x + a \hat\mu) \nnn
&& + U_\nu^\dagger(x-a \hat\nu) U_\mu(x-a \hat\nu) U_\nu(x-a\hat\nu+a\hat\mu)] \}
\eeq
where $P$ projects to a unitary matrix, which for the
SU(2) group that we work with is accomplished simply
by
\beq
A \to \frac{A}{\sqrt{\det A}}
\eeq
for a matrix $A$.  We take the smearing parameter to be $\alpha=0.5$.
We find that two smearing steps works well for yielding
the perturbatively required increase in $g^2(N,r,L)$ with $N$ at weak bare coupling.
For this reason we use two smearing steps throughout this work.

\subsection{Interpolation}
In order to obtain a value of $g^2$ for $r=1/4$
with arbitrary $N$, it is necessary
to interpolate to values of $\Rhat$ that are
not half-integral.  For this purpose Bilgici
et al.~introduce a quadratic hypothesis:
\beq
k g^2(\Rhat) \equiv \Rhat^2 \chitil(\Rhat a) \approx c_0 + c_1 \Rhat + c_2 \Rhat^2
\eeq
However, the obtained value is somewhat different if a cubic hypothesis
is chosen.  Furthermore, the result depends on which points are
included in the fit.  Here the considerations are that the smallest
values of $\Rhat$ are degraded by the smearing,
and the largest values of $\Rhat$ start to reflect
periodicity.  We have adopted the strategy of varying
all these choices and then using the variation of $kg^2$
as a measure of systematic error in the method.  The
mean value of $kg^2$ is used as our final estimate.
Further details are given in Appendix \ref{system}.

\subsection{Results}
\begin{figure}
\begin{center}
\includegraphics[width=4in,height=3in]{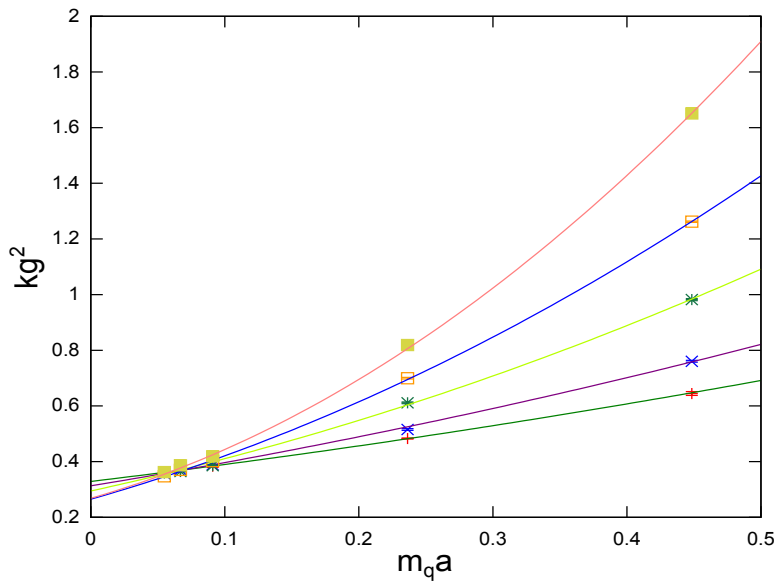}
\caption{Values of $k g^2$ for $\beta=2.25$, $r=1/4$.
At large mass, $g^2$ increases with increasing $N$, so that
the curves are ordered top to bottom, $N=24,20,16,12,10$.  However,
in the massless limit, the ordering of the curves shows a
trend of reversal, indicating backward running. \label{extrap}}
\end{center}
\end{figure}

Now that we have outlined how we obtained the values of
$k(r) g^2(N,r,L)$, we proceed to discuss our results.\footnote{We do
not need the actual value of $k(r=1/4)$ since it is a fixed
constant independent of $N$.  However, in Appendix \ref{altdef}
we consider an alternate definition of the running coupling where
$k$ depends on $N$, and in that case we do determine and divide by it.}
We have measured $kg^2$ for lattices $N=10,12,16,20,24$ with
bare coupling $\beta=2.25$.  Note that this value of $\beta$ is
at weaker coupling than the one where a bulk phase transition occurs, $\beta \approx 2$,
as was found in \cite{Catterall:2008qk}.  It is also at a stronger
coupling than the $\beta$ where there was some evidence for fixed
point behavior in the Schr\"odinger functional study \cite{Hietanen:2009az}.  
Thus this point is continuously connected to the continuum limit
through finite mass, but is expected to be on the strong side
of the infrared fixed point.  That is, we have reason to suspect
backwards running for the choice $\beta=2.25$.
We have performed our
simulations for five values of the bare mass $m_0 a$,
shown in Table \ref{masses2}.  We have measured the
PCAC mass $m_q a$ from the largest lattices ($N=24$), where
there is the least systematic error from finite volume.
These are also given in Table \ref{masses2}.  We
then extrapolate $k(r) g^2(N,r,L)$ to the $m_q=0$ limit
with quadratic and cubit fits to the data.  In the case of
the quadratic fit, we also consider the case where only
the lightest four masses are included in the fit.  
The results for the measured values of $kg^2$ are summarized
in Table~\ref{results} and the results of the
zero mass fits are presented in Table~\ref{results0}.
The results of Table \ref{results} are displayed along with the
quadratic fits to all data in Fig.~\ref{extrap}.  

At large mass one sees the behavior characteristic
of asymptotic freedom:
\beq
g^2(24) > g^2(20) > g^2(16) > g^2(12) > g^2(10).
\eeq
In the case of the quadratic fit to all data,
when we extrapolate to the chiral limit the trend
reverses:
\beq
g^2(24) \approx g^2(20) < g^2(16) < g^2(12) \approx g^2(10).
\eeq
Thus for this extrapolation we find backward running in the massless limit.
However, in the other two extrapolations of Table~\ref{results0}, what
we see is that there is no clear pattern, but instead a rough
equality:
\beq
g^2(24) \sim g^2(20) \sim g^2(16) \sim g^2(12) \sim g^2(10).
\eeq
Thus what one has in this case is evidence for an absence
of running.

We note that the $\chi^2$ per degree of freedom (d.o.f.) is
quite large for most of the fits.  The error estimates take
this into account but the implication is that the mass dependence
of $kg^2$ is complicated and not a simple polynomial.
In such a case on can question the reliability of the
zero mass extrapolations.  Appeal must finally be made
to the results at $m_0 a = -1.18$, which shows a rough
equality between the values of $kg^2$.  We expect that
this will also hold for the massless limit.  In conclusion,
our results seem to favor an absence of running, though
there is a hint of backwards running from one of the extrapolations.

The behavior that we have observed is consistent with
the existence of an infrared fixed point.  We regard our results
as suggestive that Minimal Walking Technicolor does not
actually walk, but is instead inside the conformal window.
This is supportive of the findings of Schr\"odinger
functional studies, but now by a different method.

\begin{table}
\begin{center}
\begin{tabular}{|c|c|} \hline \hline
$m_0 a$ & $m_q a$ \\ \hline \hline
 -1.000 & 0.448406(5) \\ \hline
 -1.100 & 0.236337(4) \\ \hline
 -1.165 & 0.090917(2) \\ \hline
 -1.175 & 0.066873(6) \\ \hline
 -1.180 & 0.054890(3) \\ \hline
\hline
\end{tabular}
\caption{PCAC mass obtained from $24^3 \times 48$ lattices. \label{masses2} }
\end{center}
\end{table}

\begin{table}
\begin{center}
\begin{tabular}{|c|c|c|c|c|c|} \hline \hline
	& $m_0a=-1.000$ &	$m_0a=-1.100$ & $m_0a=-1.165$	&	$m_0a=-1.175$	&	$m_0a=-1.180$	\\ \hline \hline
$10^3 \times 20$ & 0.6448(67)(0) &	0.4832(18)(0) & 0.38394(48)(0) & 0.36808(87)(0) & 0.3630(10)(0) \\ \hline
$12^3 \times 24$ & 0.7604(6)(32) & 0.5154(2)(32) & 0.3884(2)(16) & 0.37183(16)(66) & 0.35661(19)(35) \\ \hline
$16^3 \times 32$ & 0.9820(10)(37) & 0.6114(8)(31) & 0.3954(3)(18) & 0.3647(2)(13) &	0.35906(20)(97) \\ \hline
$20^3 \times 40$ & 1.2621(12)(21) & 0.6998(11)(28) & 0.4018(3)(13) & 0.3703(4)(22) & 0.3474(3)(17) \\ \hline
$24^3 \times 48$ & 1.6510(15)(7) & 0.8109(15)(12) & 0.4192(5)(10) & 0.3870(24)(7) & 0.3623(8)(18) \\
\hline \hline
\end{tabular}
\caption{Results for $kg^2$ for the various sizes of lattices and bare masses.  The first error in $kg^2$ is statistical, determined by jackknife
analysis of fits to the Creutz ratios as described in the text.  The
second error in $kg^2$ is systematic, obtained from varying the
method of interpolation used to obtain the value at $\Rhat=0.25 (L/a)$.
This is described in more detail in Appendix blah. \label{results} }
\end{center}
\end{table}

\begin{table}
\begin{center}
\begin{tabular}{|c|c|c|c|c|c|c|} \hline \hline
& \multicolumn{2}{|c|}{Quadratic, last 4 pts.} & \multicolumn{2}{c}{Quadratic, all 5 pts.}
& \multicolumn{2}{|c|}{Cubic, all 5 pts.} \\ \hline
 & $kg^2$ & $\chi^2$/d.o.f. &$kg^2$ & $\chi^2$/d.o.f. &$kg^2$ & $\chi^2$/d.o.f. \\ \hline \hline
$10^3 \times 20$ & 0.3312(49) & 2.52 &	0.3288(26) &	1.75 &	0.3321(58) &	2.44 \\ \hline
$12^3 \times 24$ & 0.297(17)	& 23.6 &	0.3130(95) &	25.7 &	0.291(20)	&21.3 \\ \hline
$16^3 \times 32$ & 0.319(20)	& 12.7 &	0.295(11) &	19.0 &	0.327(23)	&11.4 \\ \hline
$20^3 \times 40$ & 0.284(15)	& 5.04 &	0.2651(85) &	7.88&	0.288(18) &	5.36 \\ \hline
$24^3 \times 48$ & 0.313(18)	& 7.12 &	0.268(14)	& 29.5&	0.321(22) &	7.81 \\
\hline \hline
\end{tabular}
\caption{Results for the zero mass extrapolation of $kg^2$ for the various sizes of lattices.  \label{results0} }
\end{center}
\end{table}

\section{Conclusions}
We have found that the qualitative behavior of the
coupling $g^2(N,r,L)$, defined through the Creutz ratio, is consistent
with the existence of an infrared fixed point.  We arrived at this
conclusion by arguing that two different behaviors
are possible, sketched in Figs.~\ref{confine} and \ref{fixed}.
We compared our data for $g^2(N,r,L)$ to these
figures and find that as the fermion mass is reduced
toward zero, the ordering of the points either reverses
at strong bare couplings or is roughly constant, corresponding
to the region where there is a crossing
of curves in Fig.~\ref{fixed}.

While it would be disappointing from a phenomenological
point of view if Minimal Walking Technicolor is actually
conformal in the infrared, it is nevertheless significant
that by more than one method lattice field theory technique
evidence has been obtained for the existence of an infrared fixed point.
Furthermore, conformal theories are inherently interesting
and more detailed studies of the present gauge theory
are worth pursuing on the lattice.  

Two refinements to the present study could be performed,
which would hopefully yield firmer conclusions.  First,
smaller PCAC masses should be simulated so that a
stronger case for what happens in the chiral limit
can be made.  Second, larger volumes such as $N=32$
should be simulated, so that the trends with respect
to $N$ can be amplified, hopefully clarifying whether
or not backwards running actually occurs.  We are
presently pursuing these two improvements and will
report on them in the future.

\section*{Acknowledgements}
This research was supported by the Dept.~of Energy,
Office of Science, Office of High Energy Physics, 
Grant No.~DE-FG02-08ER41575.  EW was also supported by the summer 2010 REU program
in physics at RPI, funded by the National Science Foundation
award DMR 0850934 REU site at Rensselaer.
We gratefully acknowledge the sustained use of
RPI computing resources over the course of a year,
both the on-campus SUR IBM BlueGene/L rack, as well
as continuous access to 1-4 racks of the sixteen
IBM BlueGene/L's situated at the Computational Center
for Nanotechnology Innovation.

\appendix

\section{Computing methods and resources}
Simulations are performed using Hybrid Monte Carlo \cite{Duane:1987de}, which is
the standard method for dynamical fermions.  Our lattice action
is Wilson fermions and a plaquette gauge action.  Our sampling of configurations
is separated by five molecular dynamics time units, and we use 1000 samples on all
lattices excepting $N=24$, $m_0 a= -1.175, -1.180$ where we used
700 samples in each case.  (This was the most expensive part of the computation.)  
The parallel simulation and analysis code that we use
is a modification of the Columbia Physics System,
in order to incorporate SU(2) gauge group and adjoint fermions.
It was originally developed for the study of $\Ncal=1$ super-Yang-Mills \cite{Giedt:2008xm},
but has recently been used for the Minimal Walking Technicolor 
gauge theory \cite{Catterall:2008qk,Catterall:2009sb}.
All simulations were performed on IBM BlueGene/L computers
at RPI, both an on-campus rack and one to four of the sixteen racks at the
Computational Center for Nanotechnology Innovation.  Our larger
lattices were simulated using an entire rack, while our
smaller lattices used smaller partitions.  Suffice it to say,
significant computing resources were dedicated to this
project over a period of about one year.  We estimate that
7 million BlueGene/L core hours were expended for
the research summarized in this article.  Smearing to obtain
the Wilson loops that are used for our Creutz ratios was
also performed in parallel, though this analysis step consisted of
only a small fraction of our computing time.  Finally,
we computed the correlation functions necessary for
obtaining the PCAC mass using the parallel resources,
which again only consumed a small fraction of the computing time.

The remainder of the numerical analysis for this project
was performed on a single workstation.  This involved
averaging the data with a jackknife analysis of fits
in order to obtain errors in our values of $kg^2$.
Similarly, the final steps of the analysis to obtain
PCAC masses were performed with jackknife error estimation.

\section{Alternative definition}
\label{altdef}

\begin{figure}
\begin{center}
\includegraphics[width=3in,height=5in,angle=90]{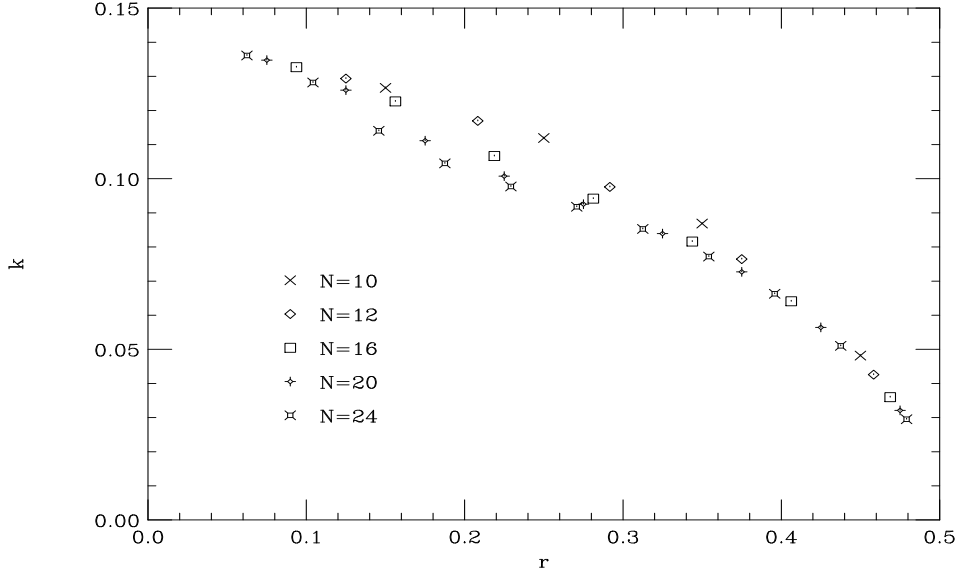}
\caption{$k(r,N)$ for the lattices that we study. \label{kfol} }
\end{center}
\end{figure}

Since there is no unique definition of the running coupling at
finite $N$, one could also use the value of $k$ at finite $N$.
That is,
\beq
g^2(N,r,L) = \frac{1}{k(r,N)} \Rhat^2 \chitil(L,a|\Rhat)
\eeq
For this, since we choose $r=1/4$ it is necessary to interpolate $k(r,N)$
between values of $r$ that are allowed for a given $N$ (we use
a quadratic fit at fixed $N$).
For the determination of $k$, we use the results of \cite{Coste:1985mn},
which leads to the following formulae.  
First their is the contribution of Wilson loops which do
not have links in the temporal direction:
\beq
c_1^{(0)}(I,J) = \frac{N_c^2-1}{N_c^2} \bigg\{ \frac{1}{4V} \sum_{p \not= 0}
\frac{1}{|P(p)|^2} \bigg( | e^{i p_1 I} - 1 |^2 \left| \sum_{k=0}^{J-1}
e^{i k p_2} \right|^2 + ( I \leftrightarrow J, 1 \leftrightarrow 2 )
\bigg) \bigg\}
\eeq
$N_c$ is the number of colors; $N_c=2$ in our case.
The quantities $|P(p)|^2$ and $V$ are defined below.
A similar formula holds for the Wilson loops with two links in
the temporal direction:
\beq
c_1^{(2)}(I,J) = \frac{N_c^2-1}{N_c^2} \bigg\{ \frac{1}{4V} \sum_{p \not= 0}
\frac{1}{|P(p)|^2} \bigg( | e^{i p_0 I} - 1 |^2 \left| \sum_{k=0}^{J-1}
e^{i k p_1} \right|^2 + ( I \leftrightarrow J, 0 \leftrightarrow 1 )
\bigg) \bigg\}
\eeq
In these expressions, the momenta take the following values,
given that there are $N$ sites in the spatial directions and $2N$ sites
in the temporal direction ($i=1,2,3$):
\beq
p_0 = n_0 \pi / N, \quad p_i = 2 n_i \pi / N, \quad n_0 = 0, \ldots, 2N-1,
\quad n_i = 0, \ldots, N-1
\eeq
Also, we have the definitions:
\beq
P_\mu(p) = e^{i p_\mu} - 1, \quad |P(p)|^2 = \sum_\mu |P_\mu(p)|^2,
\quad V = 2 N^4
\eeq
The zero-model contribution must also be included:
\beq
c_1^{(0)} = c_1(p=0) + c_1^{(0)}, \quad c_1^{(2)} = c_1(p=0) + c_1^{(2)},
\quad c_1(p=0;I,J) = \frac{N_c^2-1}{12 N_c^2} \frac{(IJ)^2}{V}
\eeq
Finally, the Creutz ratio at this order of lattice perturbation
theory is:
\beq
\chi(I,I)_{\text{1-loop}} &=& g_0^2 N_c \frac{1}{2} [ c_1^{(2)}(I,I) + c_1^{(0)}(I,I)
+ c_1^{(2)}(I-1,I-1) \ddd + c_1^{(0)}(I-1,I-1) - c_1^{(2)}(I,I-1) - c_1^{(2)}(I-1,I)
- 2 c_1^{(0)}(I,I-1) ]
\eeq
where $g_0$ is the bare coupling.  From this we extract $k$ at finite $N$:
\beq
k(r,N) = \frac{1}{g_0^2} \(I - \half\)^2 \chi(I,I)_{\text{1-loop}}, \quad
r = \(I - \half\)/N.
\eeq
In Fig.~\ref{kfol} we show $k(r,N)$ for the lattices that we study,
taking the range $I=2,\ldots,N/2$.
We note that our values of $k(r,N)$ are in agreement with
those of \cite{Bilgici:2008mt,Bilgici:2009kh}, once the difference
in color factors $(N_c^2 - 1)/N_c$ is
taken into account (they have $N_c=3$).

We have calculated $g^2$ by this alternate
approach, for the case of quadratic
extrapolation fitted to all data points,
and the results are summarized in Table \ref{kalt}.
The behavior of decreasing coupling with increasing
$N$ seems to persist with this modified definition
of the running coupling.

\begin{table}
\begin{center}
\begin{tabular}{|c|c|} \hline
$k(1/4,N)$ & $g^2$ \\  \hline
0.111952	& 2.937(23) \\
0.108871	& 2.875(87) \\
0.103207	& 2.85(10) \\
0.0995411	& 2.663(85) \\
0.0971746	& 2.76(14) \\ \hline
\end{tabular}
\end{center}
\caption{$g^2$ calculated by the method where different
values of $k$ are used for each $N$. \label{kalt} }
\end{table}

\section{Systematic error estimates}
\label{system}

As described in the main text, an interpolation of $kg^2(\Rhat)$
to the value $\Rhat = N/4$ is required, since $\chi(I,I)$
only takes values at integer $I$ and $\Rhat = I - \half$.  This
allows for various choices for the range of the fit, $I_{\text{start}}, \ldots,
I_{\text{end}}$.  One can also choose to fit with a quadratic (Fit=2) or cubic (Fit=3) 
hypothesis for $kg^2(\Rhat)$, as a function of $\Rhat$.  We have varied
these options, constrained by what is possible on the various lattices,
and have used the standard deviation of the interpolated values as
a measure of systematic error.  Because of smearing it is
important that we choose $I_{\text{start}}$ as large as possible
on a given lattice.  In Table \ref{parvar} we present the
various parameter choices that were used.
The mean of the interpolated values is
used as our estimate for $kg^2$.

\begin{table}
\begin{center}
\begin{tabular}{|c|c|c|c|}
\hline \hline
Lattice & $I_{\text{start}}$ & $I_{\text{end}}$ & Fit \\ \hline \hline
$10^3 \times 20$ & 2 & 5 & 2 \\ \hline
$12^3 \times 24$ & 3 & 7,8 & 2,3 \\ \hline
$16^3 \times 32$ & 3,4 & 7,8 & 2,3 \\ \hline
$20^3 \times 40$ & 3,4,5 & 7,8 & 2,3 \\ \hline
$24^3 \times 48$ & 4,5 & 9,10 & 2,3 \\ \hline \hline
\end{tabular}
\caption{Choices of parameters that were varied for the fit that is used in the interpolation
to obtain $k g^2(\Rhat = N/4)$. \label{parvar} }
\end{center}
\end{table}

\bibliography{crrg2}
\bibliographystyle{unsrt}

\end{document}